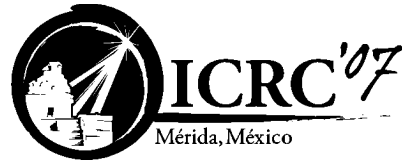

# The Cosmic Ray Observatory Project: A Statewide Outreach and Education Experiment in Nebraska

D. CLAES AND G. SNOW[1]

[1]*Department of Physics and Astronomy, University of Nebraska, Lincoln, NE 68588-0111 USA*
*dclaes@unlhep.unl.edu, gsnow@unlhep.unl.edu*

**Abstract:** The Cosmic Ray Observatory Project (CROP) is a statewide education and research experiment involving Nebraska high school students, teachers and university undergraduates in the study of extensive cosmic-ray air showers. A network of high school teams construct, install, and operate school-based detectors in coordination with University of Nebraska physics professors and graduate students. The detector system at each school is an array of scintillation counters recycled from the Chicago Air Shower Array in weather-proof enclosures on the school roof, with a GPS receiver providing a time stamp for cosmic-ray events. The detectors are connected to triggering electronics and a data-acquisition PC inside the building. Students share data via the Internet to search for time coincidences with other sites. Funded by the National Science Foundation, CROP has enlisted 29 schools with the aim of expanding to the 314 high schools in the state over several years. This report highlights both the scientific and professional development achievements of the project to date.

## Introduction

For the past 7 years, the University of Nebraska has been engaging teams of high school teachers and students in a genuine long-term cross-disciplinary research experience: studying correlations of extended cosmic ray air showers across the state of Nebraska. This paper summarizes the status, accomplishments, and future plans of CROP [1]. The early stages of CROP were documented in a 2001 ICRC proceedings contribution [2]. That report explained the scientific potential of CROP, the project's funding, the summer and academic year workshops in which the first school teams were enlisted, and the project's advisory and educational assessment strategies.

## The Schools, Teachers, and Students

Through continued summer workshops and academic year meetings, CROP has now trained 36 science instructors and over 150 students, representing 29 school teams, in the hands-on maintenance and use of their student-built cosmic ray scintillator detectors. In addition to clusters of schools in both Lincoln and Omaha, CROP now extends as far west as Scottsbluff and as far north as Springview. The locations of the participating schools, shown in Fig. 1, allow CROP to address its scientific goals on 3 distance scales. Building-sized showers (primary $E \geq 10^{15}$ eV) are recorded at individual schools with rates of several/minute. Using time coincidences among schools in populated areas, city sized showers ($E \geq 10^{19}$ eV) are recorded at much lower rates. The sparsely-spaced sites in western Nebraska allow CROP participants to investigate very long-distance correlations that would indicate bursts of high-energy primaries.

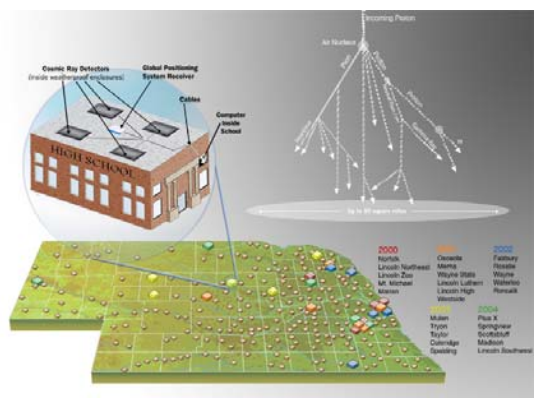

Figure 1: CROP's participating schools as of 2005 are shown as raised boxes on the map and listed at right. Remaining Nebraska schools are in brown. Inset: detector set up at a typical school.

A special one-week workshop held July 18-22, 2005, attracted students and teachers from 15 of the schools (see Fig. 2). Most of the students were new to CROP, since many of the original students had graduated. Participants brought their own equipment for refresher training and the installation of new firmware for the data acquisition card and updated LabVIEW software (see below). The last night of the workshop, all detectors were arranged in a large array and air shower data were accumulated. The workshop received press coverage in the Lincoln Journal Star and on Nebraska Public Radio and local television.

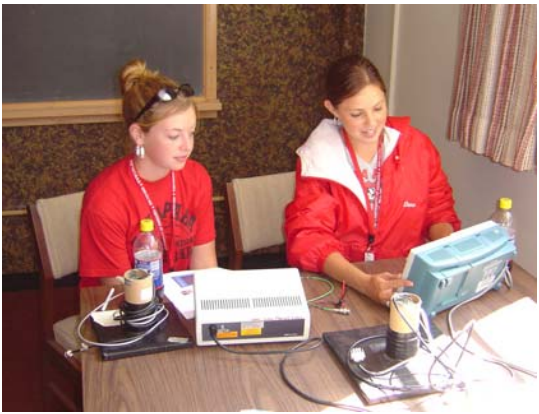

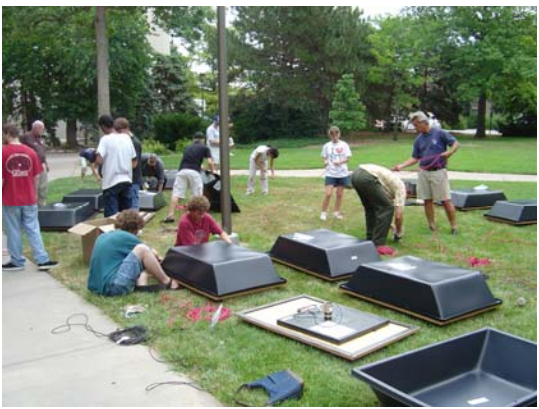

Figure 2. (Top) Students practicing oscilloscope use and (bottom) participants setting up the large detector array at the 2005 workshop.

Interest and enthusiasm from participating teachers remains high. Lincoln High School physics teacher Jim Rynearson, who joined CROP with several students in 2001, said "I use CROP to introduce students to real research. The really nice thing about CROP is that it deals with modern physics topics which most students don't see in high school." One of Rynearson's students, Ben Plowman, received a Top 5 ranking at the state Academy of Science fair based on cosmic ray measurements he made with the Lincoln High School detectors, which took him to Washington, D.C., in 2005 to compete at the national level. Another enthusiastic teacher is Fr. Michael Liebl from Mt. Michael Benedictine High School in Elkhorn, whose team joined CROP in 2000. "One of the things students learn is that real science is messy," he said. "I don't always know the answers when the students come to me." Since Mt. Michael is a boarding school, resident students have been able to compare daytime measurements of cosmic rays to nighttime when the earth blocks particles coming from the sun. The Mt. Michael team published their study [3] in a 2001 issue of The Science Teacher, the monthly publication of the National Science Teachers Association, which almost exclusively publishes articles written by university professors, high school teachers, and education specialists.

## Technical and Scientific Progress

One of CROP's planned milestones mentioned in [2] was the development of the low-cost data acquisition electronics card that serves as the interface between the scintillator detectors and the data taking PC employed in the schools. The card, described in detail in [4] and shown in Fig. 2, resulted from collaboration between physicists and engineers from CROP, WALTA [5] in Seattle, and the national QuarkNet [6] program based at Fermilab. The card has 4 threshold-adjustable discriminator inputs for the photomultiplier analog signals, a 4-channel time-to-digital converter for local coincidence and time-over-threshold measurements at 0.75 ns resolution, programmable trigger via a CPLD and microcontroller allowing 2- to 4-fold coincidence triggering, and a GPS receiver/ antenna attached via an external cable to provide event trigger time stamps at better than 100 ns accuracy.

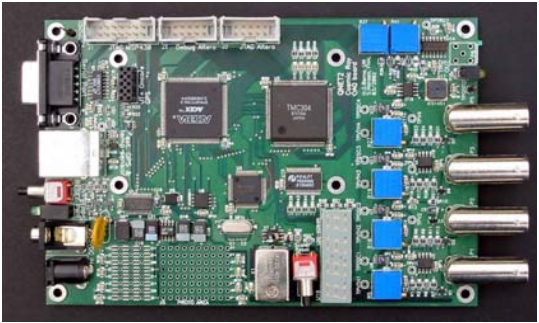

Figure 3: The data acquisition electronics card.

Onboard scalers count individual channel singles rates and triggered coincidence rates. The card outputs an ASCII event record for triggered events that is written to a file on a PC via a serial port cable. The card also has a mode in which one channel serves as a veto that is useful for muon lifetime measurements. The card has replaced the NIM electronics setups that were used in the early years of CROP for experiments performed at the schools. The project's undergraduate and graduate students have programmed a multi-purpose GUI in the LabVIEW environment that runs on the data taking PC and allows users to set trigger conditions, perform diagnostic detector tests, and control data taking run parameters.

On the analysis side, one of the project's graduate students performed a study of accidental coincidence rates as a masters degree thesis project [7] in the university's Department of Statistics. The study provides a prescription to calculate accidental rates for any combination of detectors and triggering using as input the measured singles rates for individual detectors and the time window over which one looks for coincidences. The prescription was determined from analytic calculations and verified with Monte Carlo simulations. When a typical CROP school collects air shower data, the data acquisition card is set to trigger when any 2 of the 4 rooftop detectors fire within a time window of about 1 µsec. Typical singles rates from individual detectors are of order 100 Hz. Offline analysis routines, whose inputs are the GPS time-stamped triggered event files from schools taking local data simultaneously, allow users to search for time coincidences among schools within a time window that depends on the interschool distances.

As an example, we examine the air shower data collected for a 2-week period in the summer of 2005 by 2 schools in Lincoln which are separated by about 1.5 km. We expect to observe real coincidences between these schools since they lie within the footprint of a single air shower created by a very high energy primary particle. Ten events were registered at each school within a time window of 10 µsec during this period. Applying the accidental prescription to this data reveals that the probability of observing 10 accidental coincidences is 0.083 and the probability that at least one coincidence is a real air shower is 0.916. The prescription will prove very valuable for determining the fraction of observed coincidences that may be accidental when CROP reports its scientific results.

## Phase II of CROP

Now at the end of its pilot program, CROP has established school sites in 16 the state's 19 Educational Service Units (ESUs) through which the schools are administered. For Phase II of the project, the focus will be directed toward building upon these remote sites and expanding to reach as many of the 314 high schools in Nebraska as possible. The expansion plan depends heavily on the experience and expertise of the teachers at the regional centers that have been under development since early in the program. The format of the previous 4-week or 2-week summer training workshops conducted on the university campus will be replaced by decentralized 5-day workshops held on a rotating basis at the respective ESU centers.

Two or three such remote summer workshops will be held annually, rotating among the ESUs, selected to provide maximum coverage of the state each year. Run by University of Nebraska staff and assisted by the regional CROP teachers (at least 2, enlisting help from adjacent ESUs when necessary), the workshops will introduce the equipment to be used and offer practice installing it, running the data acquisition electronics, and exercising all supporting software. New participating schools will carry fully functional detec-tors to their home institution at a given workshop's conclusion. An example of a summer's program is shown in Fig. 4. The 19 ESUs are indicated on a map of Nebraska, and 100-mile circles are drawn around 3 schools in different ESUs. Schools within a given circle will be invited to participate in that ESU's summer training workshop.

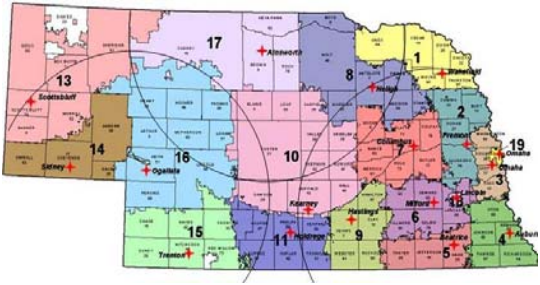

Figure 4. 100 mile circles centered on selected ESU school locations show that selecting 3 sites annually for training workshops provides access (as a day trip) to almost any school in the state.

Workshop activities will consist of (i) classroom sessions exploring topics in cosmic ray physics, charged particle detection, Monte Carlo applications, triggering, and data acquisition, and (ii) hands-on laboratory sessions during which participants will assemble and test the scintillator detectors and perform sample measurements.

Reliance on remote Web-based training and instruction modules, some currently under development, will need to be expanded considerably: video-streamed training films and presentations, regularly moderated online chat sessions, a quick turn-around hotline (phone and E-mail) staffed throughout most of the day by undergraduate and graduate students. A response plan for faulty or failing equipment must include the building and routine testing of refurbished spares. This will become an activity for later generations of students at active sites. A network of pooled spares will be developed and made available as needed by schools within each ESU. All of this will be complemented with free distribution of CDs carrying teaching materials and lesson plans.